\PassOptionsToPackage{obeyspaces,spaces}{url}
\documentclass{orgsci}

\usepackage{xcolor}
\usepackage{upquote}
\usepackage{enumitem}

\usepackage{float}        %

\usepackage{longtable}
\usepackage{pdflscape}

\providecommand{\papertablesize}{\small}

\newenvironment{threeparttable}{}{}
\newenvironment{tablenotes}[1][]{%
  \par\footnotesize\singlespacing
  \begin{list}{}{\leftmargin=0pt\labelwidth=0pt\labelsep=0pt\itemindent=0pt}%
  \RaggedRight\setlength{\parindent}{0pt}%
}{\end{list}}

\newlength{\tblmeasure}

\makeatletter
\DeclareUrlCommand\tk@repo{\urlstyle{tt}}
\DeclareUrlCommand\tk@kitfile{\urlstyle{tt}}
\DeclareUrlCommand\tk@cmd{\urlstyle{tt}}
\newcommand{\repo}{\tk@repo}
\newcommand{\kitfile}{\tk@kitfile}
\newcommand{\cmd}{\tk@cmd}
\makeatother

\newif\ifpubliclinks
\publiclinkstrue

\newcommand*{\traceossite}{trace-os.org}

\newcommand*{\traceosrepo}{github.com/amrhs-zohrevand/trace-os}

\newcommand*{\arxivid}{}

\newif\ifanonymous
\anonymousfalse
\newif\iftitlepageonly
\titlepageonlyfalse
\ifdefined\blindedbuild
  \anonymoustrue
  \publiclinksfalse
\fi
\ifdefined\titlepagebuild
  \titlepageonlytrue
\fi

\newcommand{\traceosfootnote}{%
  \ifpubliclinks
    \footnote{TRACE-OS and its documentation are available at \href{https://\traceossite}{\traceossite}.
      Its source code is at \href{https://\traceosrepo}{\traceosrepo}.%
      \ifx\arxivid\empty\else\ This paper is available at \href{https://arxiv.org/abs/\arxivid}{arXiv:\arxivid}.\fi}%
  \fi}

\begin{document}

\begin{titlepage}
\pagenumbering{Alph}
\singlespacing
{\centering
\vspace*{0.1in}
{\Large\bfseries TRACE: Governing AI-Assisted Research with\par}
\vspace{0.9em}
{\Large\bfseries Principles from Computer Science\par}

\ifanonymous\else
\vspace{1.8em}
{\large\bfseries Amirhossein (Miros) Zohrehvand\par}
\vspace{0.45em}
Leiden Institute of Advanced Computer Science (LIACS)\par
Leiden University\par
\vspace{0.35em}
\href{mailto:a.h.zohrehvand@liacs.leidenuniv.nl}{a.h.zohrehvand@liacs.leidenuniv.nl}\par

\vspace{1.2em}
{\large\bfseries Aldona Kapacinskaite\par}
\vspace{0.45em}
Strategy Area\par
INSEAD\par
\vspace{0.35em}
\href{mailto:aldona.kapacinskaite@insead.edu}{aldona.kapacinskaite@insead.edu}\par
\fi

\vspace{1.2em}
October 2026\par
}

\vspace{1.2em}
\begin{smjabstract}
Artificial intelligence (AI) has made it easier than ever to produce research outputs.
Adding an agentic tool to a research team's work potentially extends the team's capabilities but calls for an active approach to organizing the research process.
The pressing question, therefore, is how research teams should organize their work with this technology.
We contend that the question is not new: software engineering and security research have long studied how teams of people working with machines keep control of projects with many contributors.
We distill principles from the practices these literatures report and transpose them to AI-assisted research.
The principles form TRACE: Track records, Review against sources, Accept independently, Control transitions, and Evaluate corrections.
Together, the five principles give researchers a common basis for remaining in control while integrating AI into their workflows.
By bounding units of work and attending to how each is shaped, they can help researchers pace AI-assisted work to human attention rather than to the speed at which AI produces content.
We also introduce TRACE-OS, an operating system for AI-assisted research projects.
It puts the principles into practice through shared instructions, tools, and records that fit researchers' existing workflows.
With TRACE, we aim to join a conversation about the governing principles through which the research community can benefit from AI. We contend that a simple enough and useful standard can contribute to harnessing AI capabilities, allowing teams to delegate parts of their work to AI, while also building trust within teams and the broader research community.

\end{smjabstract}

\vfill
\ifanonymous\else
\smjacknowledgements{We thank Anil Doshi, Arianna Marchetti, Bart Vanneste, Joost Visser, as well as participants at the UCL SOM seminar series and the INSEAD PhD workshop for their helpful comments.}
\fi
\end{titlepage}
\pagenumbering{arabic}
\iftitlepageonly\expandafter\enddocument